\documentclass[10pt,conference]{IEEEtran}
\IEEEoverridecommandlockouts \ifx\pdfoutput\undefined
\usepackage{graphicx}
\else
\usepackage[pdftex]{graphicx}
\fi
\usepackage{cite, amsmath, amsfonts, amssymb, psfrag}
\newtheorem{definition}{{Definition}}
\newtheorem{theorem}{{Theorem}}
\newtheorem{lemma}{{Lemma}}
\newtheorem{example}{{Example}}

\begin{document}

\title{Slepian-Wolf Code Design via Source-Channel Correspondence}

\author{\authorblockN{Jun Chen}
\authorblockA{University of Illinois at Urbana-Champaign\\
Urbana, IL 61801, USA\\
Email: junchen@ifp.uiuc.edu}
\and
\authorblockN{Dake He}
\authorblockA{IBM T. J. Watson Research Center\\
Yorktown Heights, NY 10598, USA\\
Email: dakehe@us.ibm.com} \and
\authorblockN{Ashish Jagmohan}
\authorblockA{IIBM T. J. Watson Research Center\\
Yorktown Heights, NY 10598, USA\\
Email: ashishja@us.ibm.com}
 }
%

\maketitle

\begin{abstract}
We consider Slepian-Wolf code design based on LDPC (low-density
parity-check) coset codes for memoryless source-side information
pairs. A density evolution formula, equipped with a concentration
theorem, is derived for Slepian-Wolf coding based on LDPC coset
codes. As a consequence, an intimate connection between
Slepian-Wolf coding and channel coding is established.
Specifically we show that, under density evolution, design of
binary LDPC coset codes for Slepian-Wolf coding of an arbitrary
memoryless source-side information pair reduces to design of
binary LDPC codes for binary-input output-symmetric channels
without loss of optimality. With this connection, many classic
results in channel coding can be easily translated into the
Slepian-Wolf setting.
\end{abstract}

\section{Introduction}\label{intro}

Consider the problem of encoding $X$ with side information $Y$ at
the decoder. Here $(X,Y)$ are two memoryless sources with joint
probability distribution $P(x,y)$ on
$\mathcal{X}\times\mathcal{Y}$. This is a special case of
Slepian-Wolf coding. General cases can be reduced to this special
case via either time-sharing or source splitting. For this special
case, the Slepian-Wolf theorem states that the minimum rate for
reconstructing $X$ is $H(X|Y)$. For simplicity, we assume
$\mathcal{X}=\{0,1\}$ and $|\mathcal{Y}|<\infty$ throughout this
paper. The general finite-alphabet case can be reduced to this
special case via multilevel coding.  In the literature many
low-complexity Slepian-Wolf coding schemes have been proposed
\cite{PR03,CLME05,SRP04,SLXG05}, almost all of which are based on
ideas from channel coding, and simply use some binary linear
channel codes as Slepian-Wolf codes. In this approach it is
implicitly assumed that a good channel code for a channel is also
a good Slepian-Wolf code for the same channel linking the source
and the side information; yet few justifications have been
presented except for \cite{Wyner74}. Motivated by these
observations, in this paper we consider Slepian-Wolf code design
based on binary LDPC coset codes (see Section~\ref{review} for
details). To this end we derive the density evolution formula for
Slepian-Wolf coding, equipped with a concentration theorem. An
intimate connection between Slepian-Wolf coding and channel coding
is then established. Specifically we show that, under density
evolution, any Slepian-Wolf source coding problem, where the joint
distribution $P(x, y)$ can be arbitrary, is equivalent to a
channel coding problem for a binary-input output-symmetric
channel. Note that this channel is often different from the
channel between the source and the side information in the
original Slepian-Wolf coding problem. This is in sharp contrast to
the practice in the works reviewed above where the two channels
are assumed the same.

The rest of this paper is organized as follows. In Section
\ref{review}, we review some basic results in the channel coding
theory. The emphasis will be on binary linear codes with the
practical belief-propagation decoding algorithm. In Section
\ref{BP_DE_Concentration}, we develop the belief-propagation
algorithm for Slepian-Wolf coding. The associated density
evolution formula and concentration theorem are also provided. An
intimate connection between Slepian-Wolf coding and channel coding
under density evolution is established in Section
\ref{SCconnection}.  We conclude the paper in Section
\ref{Conclusion}.

\section{Review of Channel Coding Theory}\label{review}

Any binary linear code $\mathcal{C}=\{\mathbf{c}\}$ can be
expressed as the set of solutions $\mathbf{c}$ to a parity check
equation $H\mathbf{c}=\mathbf{0}$, where $H$ is called the parity
check matrix of $\mathcal{C}$, and here multiplication and
addition are modulo 2. Given some general syndrome
$\mathbf{s}\in\{0,1\}^{n-k}$, the set of all $n$-length vectors
$\mathbf{x}$ satisfying $H\mathbf{x}=\mathbf{s}$ is called a coset
$\mathcal{C}_{\mathbf{s}}$. A $(d_v,d_c)$-regular
low-density-parity-check (LDPC) code is a binary linear code
determined by the condition that every codeword bit participates
in exactly $d_v$ parity-check equations and that every such check
equation involves exactly $d_c$ codeword bits. Given the parity
check matrix $H$, we can construct a bipartite graph with $n$
variable nodes and $m=n-k$ check nodes. Each variable node
corresponds to one bit of the codeword, and each check node
corresponds to one parity-check equation. Edges in the graph
connect variable nodes to check nodes and are in one-to-one
correspondence with the nonzero entries of $H$. The ensemble
$\mathcal{C}^n(d_v,d_c)$ of $(d_v,d_c)$-regular LDPC codes of
length $n$ is defined in \cite{LMSSS97}. We can also define the
irregular code ensemble $\mathcal{C}^n(\lambda,\rho)$, where
$(\lambda,\rho)$ denotes a degree distribution pair
\cite{LMSSS97}.

The belief-propagation algorithm is an iterative message-passing
algorithm. Let $m^{(l)}_{vc}$ denote the message sent from
variable node $v$ to its incident check node $c$ in the $l$th
iteration. Similarly, let $m^{(l)}_{cv}$ denote the message sent
from check node $c$ to its incident variable node $v$ in the $l$th
iteration. The update equations for the messages under belief
propagation are described below:
\begin{eqnarray*}
m^{(l)}_{vc}=\left\{\begin{array}{ll}
  m_0,  &\mbox{if } l=0\\
  m_0+\sum\limits_{c'\in C_v\setminus\{c\}}m^{(l)}_{c'v},  \quad\quad &\mbox{if } l\geq 1\\
\end{array}\right.&
\end{eqnarray*}
\begin{eqnarray*}
\hspace{-0.58in}m^{(l)}_{cv}=\gamma^{-1}\left(\sum\limits_{v'\in
V_c\setminus\{v\}}\gamma\left(m^{(l-1)}_{v'c}\right)\right).
\end{eqnarray*}
where $C_v$ is the set of check nodes incident to variable node
$v$, $V_c$ is the set of variable nodes incident to check node
$c$, $m_0\triangleq\ln\frac{P(y_i|x_i=0)}{P(y_i|x_i=1)}$ is the
initial message associated with the variable node $v$, and
$\gamma(x)=\left(\textrm{sgn}(x),-\ln\tanh\left|\frac{x}{2}\right|\right)$.

The performance of LDPC codes under the belief-propagation
algorithm is relatively well-understood for binary-input
output-symmetric (BIOS) channels.
\begin{definition}
A binary input channel with transition probability function
$p(\cdot|\cdot): {\cal X} \times {\cal Y} \to [0, 1]$ from ${\cal
X}$ to ${\cal Y}$ is output-symmetric if there exists an injective
map $\varphi:\mathcal{Y}\rightarrow\mathbb{R}$ such that
\begin{eqnarray*}
p(\varphi(y_t)=\varphi(y)|x_t=0)=p(\varphi(y_t)=-\varphi(y)|x_t=1),\forall
y\in\mathcal{Y},
\end{eqnarray*}
where $y_t$ is the channel output in response to channel input
$x_t$.
\end{definition}

An important property of the BIOS channel is that under the
belief-propagation algorithm, the decoding error probability is
independent of the transmitted codeword. So without loss of
generality, we can assume the all-zero codeword is transmitted.

In order to analyze the asymptotic (in codeword length)
performance of the LDPC code ensemble
$\mathcal{C}^n(\lambda,\rho)$, a powerful technique called density
evolution is developed in \cite{RU01,RU01_2}. The iterative
density evolution formula for BIOS channels is given in the
following theorem.
\begin{theorem}[\cite{RU01_2}, Theorem 2]
For a given BIOS memoryless channel let $P^{(0)}$ denote the
initial message density of log-likelihood ratios, assuming that
the all-zero codeword was transmitted. If, for a fixed degree
distribution pair $(\lambda,\rho)$, $P^{(l)}$ denotes the density
of the messages passed from the variable nodes to the check nodes
at the $l$th iteration of belief propagation then, under the
independence assumption
\begin{eqnarray}
P^{(l)}=P^{(0)}\otimes\lambda(\Gamma^{-1}(\rho(\Gamma(P^{(l-1)}))))\label{channelDE}
\end{eqnarray}
where $\Gamma$ is the density transformation operator induced by
$\gamma$.
\end{theorem}
The theorem below provides a theoretical foundation of density
evolution.
\begin{theorem}[\cite{RU01}, Theorem 2]
Over the probability space of all graphs $\mathcal{C}^n(d_v,d_c)$
and channel realizations let $Z$ be the number of incorrect
messages among all $nd_v$ variable-to-check node messages passed
at iteration $l$. Let $p^{(l)}_e$ be the expected number of
incorrect messages passed along an edge with a tree-like directed
neighborhood of depth at least $2l$ at the $l$th iteration, i.e.,
\begin{eqnarray*}
p^{(l)}_e=\int_{-\infty}^{0^{-}}P^{(l)}(dm)+\frac{1}{2}\int_{0^{-}}^{0^{+}}P^{(l)}(dm).
\end{eqnarray*}
Then, there exist positive constants $\beta=\beta(d_v,d_c,l)$ and
$\gamma=\gamma(d_v,d_c,l)$ such that for any $\epsilon>0$ and
$n>\frac{2\gamma}{\epsilon}$ we have
\begin{eqnarray*}
\mbox{Pr}\{|Z-nd_vp^{(l)}_e|>nd_v\epsilon\}\leq
2e^{-\beta\epsilon^2n}.
\end{eqnarray*}
\end{theorem}

The main contribution of this paper is that we prove a similar
density evolution formula with an associated concentration theorem
for Slepian-Wolf coding, and establish an intimate connection
between Slepian-Wolf coding and channel coding under density
evolution.

For a given joint distribution $P(x,y)$ on
$\mathcal{X}\times\mathcal{Y}$, we can define two channels: one
from $X$ to $Y$ with transition probability $P(y|x)$, the other
from $Y$ to $X$ with transition probability $P(x|y)$. Both
channels are somehow related to Slepian-Wolf coding as we
shall discuss in the following two examples:

\begin{example}[Channel from $Y$ to $X$]
Suppose $X=Y\oplus Z$, where $X, Y$ and $Z$ all assume values in
$\{0,1\}$, $\oplus$ is the modulo-2 addition, and
$Z\sim\mbox{Ber}(q)$ (with $0\leq q\leq\frac{1}{2}$) is
independent of $Y$. Let $H$ be an $(n,k)$ binary parity-check
matrix. Let $\mathcal{C}$ (i.e., $\mathcal{C}_0$) be the linear
code with the parity check matrix $H$. Assuming that all rows of
$H$ are linearly independent, there are $2^k$ codewords in
$\mathcal{C}$, so the code rate is $(\log|\mathcal{C}|)/n=k/n$.

The following scheme was suggested by Wyner \cite{Wyner74}. Given
$\mathbf{x}$, the encoder sends the syndrome
$\mathbf{s}=H\mathbf{x}$ to the decoder. With the side information
$\mathbf{y}$, the decoder can compute
$H\mathbf{z}=H\mathbf{x}\oplus H\mathbf{y}=\mathbf{s}\oplus
H\mathbf{y}$. Syndrome decoding can be implemented to find
the minimum weight $\hat{\mathbf{z}}$ such that
$H\hat{\mathbf{z}}=H\mathbf{z}$. The decoder then claims that
$\hat{\mathbf{x}}=\mathbf{y}+\hat{\mathbf{z}}$ is the target
sequence $\mathbf{x}$. It can be shown that if the error
probability of $\mathcal{C}$ when used over the channel from $Y$ to
$X$ (i.e., a binary symmetric channel with crossover probability
$q$) under syndrome decoding is $\epsilon$, then the above
Slepian-Wolf coding scheme is also of error probability
$\epsilon$. Furthermore, if $\mathcal{C}$ is capacity-achieving
for the channel from $Y$ to $X$, i.e, the rate of $\mathcal{C}$ is
$1-H(q)$, then the rate of this Slepian-Wolf coding scheme is
$H(q)$, which is exactly the Slepian-Wolf limit.
\end{example}

\begin{example}[Channel from $X$ to $Y$]
Suppose $X$ is uniformly distributed over $\{0,1\}$ and $P(y|x)$
is a BIOS channel.

The encoding procedure is the same as that in Example 1.  Given
$\mathbf{x}$, the encoder finds the coset $C_{\mathbf{s}}$ that
contains $\mathbf{x}$ and send the syndrome $\mathbf{s}$ to the
decoder. So the encoder rate is $(n-k)/n$. Given the side
information $\mathbf{y}$, the decoder tries to recover
$\mathbf{x}$ using the belief propagation decoding algorithm for
$C_{\mathbf{s}}$. For a BIOS channel, the decoding error
probability is the same for all coset codes
$\mathcal{C}_{\mathbf{s}}$ under the belief-propagation algorithm.
So if $\mathcal{C}_{\mathbf{0}}$ is a linear code for channel
$P(y|x)$ with error probability $\epsilon$ under
belief-propagation decoding, then the error probability of the
above Slepian-Wolf coset coding scheme is also $\epsilon$.
Furthermore, assuming $\mathcal{C}_{\mathbf{0}}$ is a capacity
achieving linear code for channel $P(y|x)$, the above coding
scheme is then of rate $1-C=H(X|Y)$, which is exactly the
Slepian-Wolf limit. Here $C$ is the capacity of channel $P(y|x)$.

When $X$ is nonuniform, we can still use the above coset coding
scheme as long as $\mathcal{C}_{\mathbf{0}}$ is a good channel
code for channel $P(y|x)$ under the belief-propagation decoding
algorithm. The reason is that for a BIOS channel, the error
probability resulting from the belief propagation decoding
algorithm is the same for every codeword in every coset.
Nonetheless, since $1-C
>H(X|Y)$ when $X$ is nonuniform, we see that the above coset
coding scheme fails to achieve the Slepian-Wolf limit even when
$\mathcal{C}_{\mathbf{0}}$ is an optimal channel code for channel
$P(x|y)$. This phenomenon has been observed in \cite{LTB04}.

When $P(y|x)$ is not output-symmetric, the decoding error
probability, under the belief-propagation algorithm, is in general
different for different codewords in each coset, and also
different for different cosets. In this case, the connection
between channel coding for channel $P(y|x)$ and Slepian-Wolf
coding is not clear.
\end{example}

We have seen that although the above two examples exhibit some
interesting connections between channel coding (either for the
channel from $X$ to $Y$ or the channel from $Y$ to $X$) and
Slepian-Wolf coding, both of them have severe limitations. In this
paper we shall provide a general framework, which includes these
two examples as special cases, and within the framework establish
the connection between channel coding and Slepian-Wolf coding. It
should be emphasized that in our framework, $P(x)$ does not need
to be uniform, and $P(y|x)$ does not need to be output-symmetric.

\section{Belief-Propagation Algorithm, Density Evolution and Concentration
Theorem}\label{BP_DE_Concentration}

We use the same encoding method as that in Example 1 and Example
2. We first fix a parity check matrix $H$. Given $\mathbf{x}$, the
encoder sends the syndrome $\mathbf{s}=H\mathbf{x}$ to the
decoder. But we do not use the channel decoding method in Example
2. The reason is that in channel coding, codewords are assumed to
be equally probable, but in Slepian-Wolf coding, codewords are
generated by $P(x)$, and are in general not equally probable if
$X$ is not uniform over
$\{0,1\}$. 
It turns out that it is easy to incorporate the prior distribution
$P(x)$ into the belief-propagation. The update equations for the
messages in this Slepian-Wolf decoding are described below:
\begin{eqnarray*}
m^{(l)}_{vc}=\left\{\begin{array}{ll}
  m_0,  &\mbox{if } l=0\\
  m_0+\sum\limits_{c'\in C_v\setminus\{c\}}m^{(l)}_{c'v},  \quad\quad &\mbox{if } l\geq 1\\
\end{array}\right.&
\end{eqnarray*}
\begin{eqnarray}
\hspace{-0.25in}m^{(l)}_{cv}=(-1)^s\gamma^{-1}\left(\sum\limits_{v'\in
V_c\setminus\{v\}}\gamma\left(m^{(l-1)}_{v'c}\right)\right).\label{checknodeop}
\end{eqnarray}
where
$m_0\triangleq\ln\frac{P(x_i=0|y_i)}{P(x_i=1|y_i)}=\ln\frac{P(x_i=0,y_i)}{P(x_i=1,y_i)}$,
and $s$ is the syndrome value associated with check node $c$. It
can be verified that this algorithm produces the exact
symbol-by-symbol \textit{a posteriori} estimation of $\mathbf{x}$
given $\mathbf{y}$ when the underlying bipartite graph is a tree.
We can see that the only difference from the channel decoding case
is the definition of initial message $m_0$. This decoding scheme
can be viewed as a MAP-version of the belief propagation
algorithm, while the channel decoding scheme can be viewed as a
ML-version of the belief propagation algorithm.

Now we proceed to develop the density evolution formula for this
belief-propagation algorithm. We use the standard tree assumption.
Let $P^{(l)}(x)$ $(x=0,1)$ be the message distribution from a
variable node to a check node at the $l$th iteration conditioned
on the event that the variable value is $x$. Similarly, let $Q^{(l)}(x)$
$(x=0,1)$ be the message distribution from a check node to a
variable node at the $l$th iteration conditioned on the event that the
target variable value is $x$. Assume $P(X=0)=p$. Let $\langle
P^{(l)}\rangle=pP^{(l)}(0)+(1-p)P^{(l)}(1)\circ I^{-1}$, where
$I(m)\triangleq -m$ is a parity reversing function. We have
\begin{eqnarray}
p^{(l)}_e=\int_{-\infty}^{0^{-}}\langle P^{(l)}\rangle
(dm)+\frac{1}{2}\int_{0^{-}}^{0^{+}}\langle P^{(l)}\rangle
(dm),\label{errorprob}
\end{eqnarray}
where $p^{(l)}_e$ is the expected number of incorrect messages
sent from a variable node at the $l$th iteration.

The following theorem provides a density-evolution formula for
$\langle P^{(l)}\rangle$.
\begin{theorem}\label{theorem3}
Under the tree assumption,
\begin{eqnarray}
\langle P^{(l)}\rangle=\langle
P^{(0)}\rangle\otimes\lambda(\Gamma^{-1}(\rho(\Gamma(\langle
P^{(l-1)}\rangle)))). \label{sourceDE}
\end{eqnarray}
\end{theorem}

{\em Remark}: Theorem \ref{theorem3} does not directly follow from
the approach in \cite{RU01_2} since the all-zero codeword
assumption is not valid in our setting.

\begin{theorem}[Concentration Theorem]\label{thm3}
Over the probability space of all graphs $\mathcal{C}^n(d_v,d_c)$
and source realizations let $Z$ be the number of incorrect
messages among all $nd_v$ variable-to-check node messages passed
at iteration $l$. Then, there exist positive constants
$\beta=\beta(d_v,d_c,l)$ and $\gamma=\gamma(d_v,d_c,l)$ such that
for any $\epsilon>0$ and $n>\frac{2\gamma}{\epsilon}$ we have
\begin{eqnarray*}
\mbox{Pr}\{|Z-nd_vp^{(l)}_e|>nd_v\epsilon\}\leq
2e^{-\beta\epsilon^2n}.
\end{eqnarray*}
\end{theorem}

It can be verified that the density evolution formula
(\ref{sourceDE}) and concentration theorem (Theorem \ref{thm3})
do not depend on the definition of the initial message, i.e., they
still hold if we replace $\langle P^{(0)}\rangle$ by an arbitrary
probability distribution. This provides us a useful tool to study
the problem of distribution mismatch. In many applications, the
true source distribution cannot be estimated perfectly. For
example, suppose $P(x,y)$ is the true source distribution and
$P_{es}(x,y)$ is the estimated source distribution. The initial
message is then given by
$m_0=\log\frac{P_{es}(x=0|y)}{P_{es}(x=1|y)}$.

Let $P^{(l)}_{es}(x)$ $(x=0,1)$ be the density of the message from a
variable node to a check node at the $l$th iteration conditioned
on the event that the variable value is $x$. Let $\langle
P^{(l)}_{es}\rangle=P(x=0)P^{(l)}_{es}(0)+P(x=1)P^{(l)}_{es}(1)\circ
I^{-1}$. The density evolution formula of this mismatch problem is
\begin{eqnarray}
\langle P^{(l)}_{es}\rangle=\langle
P^{(0)}_{es}\rangle\otimes\lambda(\Gamma^{-1}(\rho(\Gamma(\langle
P^{(l-1)}_{es}\rangle)))).   \label{DEmismatch}
\end{eqnarray}
We can use the density evolution formula (\ref{DEmismatch}) to
check whether the error probability goes to zero when the
distribution mismatch occurs.

We have seen from Example 2 in Section \ref{review} that under the ML-version of the belief-propagation decoding algorithm, the
Slepian-Wolf coset coding scheme still works  even if $X$ is
nonuniform, as long as $\mathcal{C}_{\mathbf{0}}$ is a good
channel code for channel $P(y|x)$ under the belief-propagation
decoding algorithm. Actually, using the ML-version of the
belief-propagation algorithm for decoding nonuniform $X$ can be
viewed as a special case of distribution mismatch, where
\begin{eqnarray*}
\log\frac{P_{es}(x=0|y)}{P_{es}(x=1|y)}=\log\frac{P(y|x=0)}{P(y|x=1)},\quad\forall
y\in\mathcal{Y}.
\end{eqnarray*}
Thus in this example distribution mismatch does not imply decoding
failure, but it may cause rate loss.

\section{Source-Channel Correspondence}\label{SCconnection}

In the density evolution formula (\ref{channelDE}) in channel
coding and density evolution formula (\ref{sourceDE}) in
Slepian-Wolf coding, the channel and source statistics come in
only through the initial message distribution; all the remaining
operations depend only on the degree distribution. So for a fixed
degree distribution pair $(\lambda,\rho)$, if $P^{(0)}=\langle
P^{(0)}\rangle$, then the two density evolutions are completely
identical, i.e., we have $P^{(l)}=\langle P^{(l)}\rangle$ for all
$l$. So a natural question is: For a given Slepian-Wolf initial
message distribution $\langle P^{(0)}\rangle$, does there exist a
BIOS channel whose initial message distribution $P^{(0)}$ is the
same as $\langle P^{(0)}\rangle$? We now proceed to answer this
question.
\begin{definition}[\cite{RU01_2}, Definition 1]\label{symmetric}
We call a distribution $Q$ symmetric if $\int h(m)Q(dm)=\int
e^{-m}h(-m)Q(dm)$ for any function $h$ for which the integral
exists.
\end{definition}


\begin{lemma}\label{symmetricini}
$\langle P^{(l)}\rangle$ is symmetric.
\end{lemma}

{\em Remark}: The reason why $\langle P^{(l)}\rangle$ is symmetric
even when there is no symmetry in the source distribution $P(x,y)$
is that the coset coding scheme is used, and the prior
distribution $P(x)$ is incorporated in the decoding.

The following theorem is the main result of this section, which
essentially says that under belief propagation decoding, for all
$(X,Y)$ pairs, Slepian-Wolf code design with linear codes reduces
to design of codes for certain symmetric channels.

\begin{theorem}
For any source distribution $P(x,y)$ on
$\mathcal{X}\times\mathcal{Y}$ ($\mathcal{X}=\{0,1\},
|\mathcal{Y}|<\infty$) with conditional entropy $H(X|Y)$, there
exists a unique BIOS channel $P'(y'|x')$ with capacity $C$ such
that its initial message distribution $P^{(0)}$ is the same as the
initial message distribution $\langle P^{(0)}\rangle$ induced by
$P(x,y)$. Furthermore, we have $H(X|Y)+C=1$. The conversion from
$P(x,y)$ to $P'(y'|x')$ is given in Fig. 1.
\end{theorem}

{\em Remark}: It can be seen from Fig. 1 that $P'(y'|x')$ is in
general different from $P(y|x)$, and $P'(y'|x')$ is symmetric even
when $P(y|x)$ is not. Furthermore, if $P(x)$ is nonuniform, then
$P'(y'|x')$ is different from $P(y|x)$ even when $P(y|x)$ is
symmetric.

\begin{figure}[hbt]
\centering
\begin{psfrags}
\psfrag{p0}[r]{$p_0$}%
\psfrag{p1}[r]{$p_1$}%
\psfrag{q00}[c]{$q_{0,0}$}%
\psfrag{q10}[c]{$q_{1,0}$}%
\psfrag{q01}[c]{$q_{0,1}$}%
\psfrag{q11}[c]{$q_{1,1}$}%
\psfrag{q0n-2}[c]{$q_{0,n-2}$}%
\psfrag{q1n-2}[c]{$q_{1,n-2}$}%
\psfrag{q0n-1}[c]{$q_{0,n-1}$}%
\psfrag{q1n-1}[c]{$q_{1,n-1}$}%
\psfrag{p0q00}[c]{$p_0q_{0,0}$}%
\psfrag{p1q10}[c]{$p_1q_{1,0}$}%
\psfrag{p0q01}[c]{$p_0q_{0,1}$}%
\psfrag{p1q11}[c]{$p_1q_{1,1}$}%
\psfrag{p0q0n-1}[c]{$p_0q_{0,n-1}$}%
\psfrag{p1q1n-1}[c]{$p_1q_{1,n-1}$}%
\psfrag{dots}[c]{\Large$\cdots$} %
\psfrag{vdots}[c]{\Large$\vdots$} %
\includegraphics[scale=0.5]{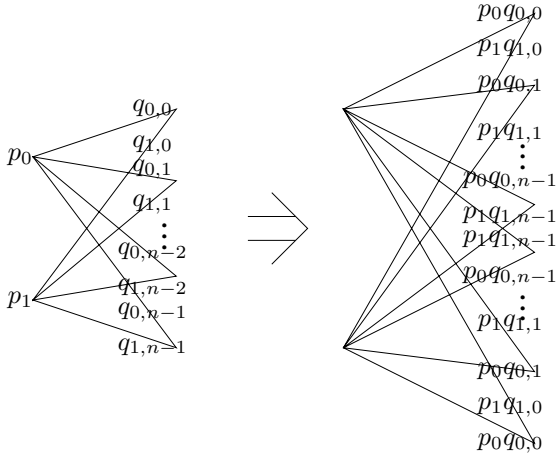}
\caption{Source-to-channel conversion}
\end{psfrags}
\end{figure}

Let $\mbox{Ch}(\cdot)$ be the function that maps $P(x,y)$ to
$P'(y'|x')$. It turns out this function is not invertible.
\begin{definition}[Equivalence]\label{equivalence}
Two sources distributions, $P(x,y)$ and $P'(x',y')$, are
equivalent if they induce the same initial message distribution
$\langle P^{(0)}\rangle$ (or if
$\mbox{Ch}(P(x,y))=\mbox{Ch}(P'(x',y'))$).
\end{definition}

For a symmetric distribution $Q$ given by
$Q\left(\ln\frac{a_i}{a_{n-1-i}}\right)=a_i$, $i=0,1,\cdots,n-1$,
where $a_i\in(0,1]$ and $\sum_{i=0}^{n-1}a_i=1$, we can compute
all source distributions for which the induced initial message
distribution $\langle P^{(0)}\rangle$ is equal to $Q$. These can be
written (possibly after relabelling) in the following parametric
form:
\begin{eqnarray}
P(y=i)&=&\alpha_i(a_i+a_{n-1-i}),\label{ydistribution}\\
P(x=0|y=i)&=&\frac{a_i}{a_i+a_{n-1-i}},\label{invreverse1}\\
P(x=1|y=i)&=&\frac{a_{n-1-i}}{a_i+a_{n-1-i}},
i=0,\cdots,n-1.\label{invreverse2}
\end{eqnarray}
where $\alpha_i\in [0,1]$, $1-\alpha_i=\alpha_{n-1-i}$,
$i=0,1,\cdots,n-1$. So for a fixed initial message distribution
with $n$ probability mass points, the set of equivalent source
distributions has totally $\lfloor\frac{n}{2}\rfloor$ degrees of
freedom. This should be contrasted with the $2n-2$ degrees of freedom
for source distributions (over $\mathcal{X}\times\mathcal{Y}$ with
$|\mathcal{X}|=2$ and $|\mathcal{Y}|=n$) under the conditional
entropy constraint. We can also see that the equivalent source
distributions must have the same reverse channel $P(x|y)$; the
freedom comes only from $P(y)$.

\begin{example}
Consider the following class of source distributions
$P(y=e)=\epsilon$, $P(y\in\{0,1\})=1-\epsilon$,
$P(x=i|y=i)=1$, $i=0,1$, $P(x=0|y=e)=P(x=1|y=e)=\frac{1}{2}$. It
is easy to verify that this class of source distributions is an
equivalence class. In fact, for any distribution $P(x,y)$ in this class,
$\mbox{Ch}(P(x,y))$ is the binary erasure channel with erasure
probability $\epsilon$. Since the capacity of the binary erasure
channel can be achieved with LDPC codes under the
belief-propagation decoding algorithm, it follows that for
distributions in this class, the Slepian-Wolf limit is achievable
with the LDPC coset coding scheme under belief-propagation
decoding.
\end{example}

\begin{example}
It can be verified that for fixed $q$, the source distributions in
Example 1 are all equivalent.
\end{example}

Given two channels, $P(y|x)$ and $P'(y'|x')$, we say
$P'(y'|x')\preceq P(y|x)$ if $P'(y'|x')$ is physically degraded
with respect to $P(y|x)$. We now generalize this concept to source
distributions.
\begin{definition}[Monotonicity]\label{monotonicity}
Given two source distributions, $P(x,y)$ and $P'(x',y')$, we say
$P'(x',y')\preceq P(x,y)$ if
$\mbox{Ch}(P'(x',y'))\preceq\mbox{Ch}(P(x,y))$.
\end{definition}

{\em Remark}: One may tend to define ``monotonicity" in the
following way: $P'(x',y')\preceq P(x,y)$ if (possibly after
relabelling) $P'(x)=P(x)$ for all $x\in\mathcal{X}$ and
$P'(y'|x')$ is physically degraded with respect to $P(y|x)$. It
turns out the ``monotonicity" in this sense also satisfies the
condition $\mbox{Ch}(P'(x',y'))\preceq\mbox{Ch}(P(x,y))$, as
illustrated in Fig. 2.

\begin{figure}[hbt]
\centering
\begin{psfrags}
\psfrag{p0}[r]{$p_0$}%
\psfrag{p1}[r]{$p_1$}%
\psfrag{q00}[c]{$q_{0,0}$}%
\psfrag{q10}[c]{$q_{1,0}$}%
\psfrag{q01}[c]{$q_{0,1}$}%
\psfrag{q11}[c]{$q_{1,1}$}%
\psfrag{q0n-2}[c]{$q_{0,n-2}$}%
\psfrag{q1n-2}[c]{$q_{1,n-2}$}%
\psfrag{q0n-1}[c]{$q_{0,n-1}$}%
\psfrag{q1n-1}[c]{$q_{1,n-1}$}%
\psfrag{p0q00}[c]{$p_0q_{0,0}$}%
\psfrag{p1q10}[c]{$p_1q_{1,0}$}%
\psfrag{p0q01}[c]{$p_0q_{0,1}$}%
\psfrag{p1q11}[c]{$p_1q_{1,1}$}%
\psfrag{p0q0n-1}[c]{$p_0q_{0,n-1}$}%
\psfrag{p1q1n-1}[c]{$p_1q_{1,n-1}$}%
\psfrag{s00}[c]{$s_{0,0}$}%
\psfrag{s10}[c]{$s_{1,0}$}%
\psfrag{s0m-1}[c]{$s_{0,m-1}$}%
\psfrag{s1m-1}[c]{$s_{1,m-1}$}%
\psfrag{sn-20}[c]{$s_{n-2,0}$}%
\psfrag{sn-10}[c]{$s_{n-1,0}$}%
\psfrag{sn-2m-1}[c]{$s_{n-2,m-1}$}%
\psfrag{sn-1m-1}[c]{$s_{n-1,m-1}$}%
\psfrag{p(x,y)}[c]{$P(x,y)$} %
\psfrag{p'(x',y')}[c]{$P'(x',y')$} %
\psfrag{ch(p(x,y))}[c]{Ch($P(x,y)$)} %
\psfrag{ch(p'(x',y'))}[c]{Ch($P'(x',y')$)} %
\psfrag{vdots}[c]{\Large$\vdots$} %
\includegraphics[scale=0.4]{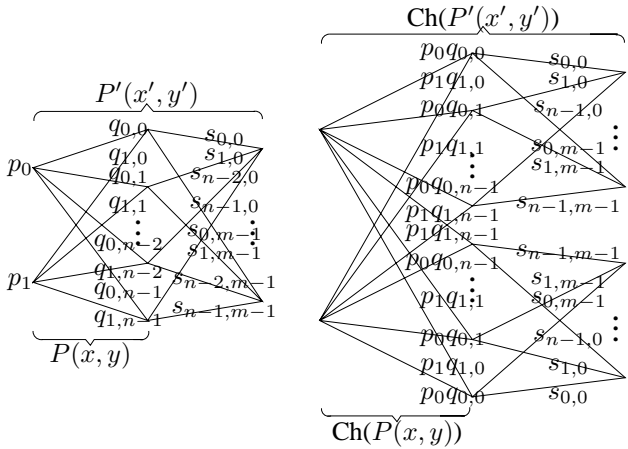}
\caption{Preservation of source monotonicity in the channel
domain}
\end{psfrags}
\end{figure}

The following theorem follows immediately from the monotonicity of
density evolution with respect to physically degraded channels.
\begin{theorem}
Suppose $P'(x',y')\preceq P(x,y)$. For any fixed degree
distribution pair $(\lambda,\rho)$, if, for $P'(x',y')$,  the
error probability under density evolution goes to zero, then it
must also go to zero for $P(x,y)$.
\end{theorem}

\begin{definition}
For a fixed degree distribution pair $(\lambda,\rho)$ and a class
of distributions $\mathcal{P}$ over
$\mathcal{X}\times\mathcal{Y}$, the feasible domain
$\mathcal{F}\subseteq\mathcal{P}$ with respect to $(\lambda,\rho)$
is the set of distributions $P(x,y)\in\mathcal{P}$ for which the
error probability under density evolution goes to zero.
\end{definition}

\begin{example}
Let $\mathcal{P}$ be the class of distributions $P(x,y)$ over
$\{0,1\}\times\{0,1\}$ such that $P(y=1|x=0)=P(y=0|x=1)$. Let
$P(x=0)=p$ and $P(y=1|x=0)=P(y=0|x=1)=q$. Then $\mathcal{P}$ can
be parameterized by $p$ and $q$ with $p,q\in[0,\frac{1}{2}]$.

Let $\mathcal{S}(R)=\left\{(p,q):H(X|Y)|_{p,q}\leq R, p, q
\in\left[0,\frac{1}{2}\right]\right\}$. For any degree
distribution pair $(\lambda,\rho)$ with syndrome rate less than or
equal to $R$, by Slepian-Wolf theorem we must have
$\mathcal{F}\subseteq\mathcal{S}(R)$. In Fig. 3 the area below the
solid curve is $\mathcal{S}(\frac{1}{2})$. We also plot the
feasible domains of several rate one-half codes.
\begin{enumerate}
\item Code 1: Its degree distribution pair is given in
\cite[Example 2]{RU01_2}. This code is designed for a binary
symmetric channel, which corresponds to $p=\frac{1}{2}$ in Fig. 3.
The feasible domain of this degree distribution pair is the area
below the ``$\circ$" curve.
\item Code 2:
$\lambda(x)=0.234029x+0.212425x^2+0.146898x^5+0.102840x^6+0.303808x^{19}$,
$\rho(x)=0.71875x^{7}+0.28125x^{8}$. This code is designed for a
binary-input AWGN channel \cite{Chung00}. The feasible domain is
the area below the ``+" curve.
\item Code 3 is the $(3,6)$-regular code. Its feasible domain is
the area below the ``*" curve.
\item Code 4 is the $(4,8)$-regular code. Its feasible domain is
the area below the ``$\square$" curve.
\end{enumerate}
It can be seen that although code 1 is designed for the case
$p=\frac{1}{2}$, it performs very well over the whole range; the
performance of code 2 is also quite good, although it is designed
for binary-input AWGN channel. This should not be too surprising
since under density evolution, every Slepian-Wolf coding problem
is equivalent to a channel coding problem for a corresponding BIOS
channel, and it is a well-known phenomenon that a code good for
one BIOS channel is likely to be good for many other BIOS
channels. Therefore, Fig. 3 is simply a manifestation of this
phenomenon in the Slepian-Wolf source coding scenario.
\end{example}

\begin{figure}[hbt]
\centering
\includegraphics[scale=0.43]{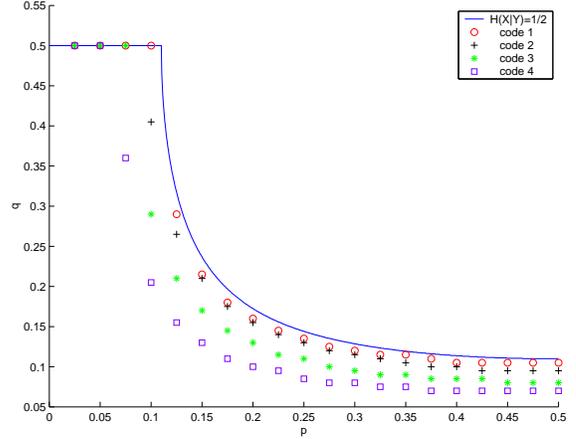}
\caption{Feasible domain}
\end{figure}

\section{Conclusion}\label{Conclusion}

We have established an intimate connection between Slepian-Wolf
coding and channel coding, which clarifies a misconception in the
area of Slepian-Wolf code design. Interested readers may refer to
\cite{CHJ06} for more details.


\end{document}